\documentclass[aps,prl,twocolumn,superscriptaddress]{revtex4}
\usepackage{amssymb}
\usepackage{amsmath}
\usepackage{graphicx}
\usepackage{epsfig}
\usepackage{amsfonts}

\begin{document}

\title{Mimicking interacting relativistic theories with stationary pulses of light}
\date{\today}
\author{Dimitris G. Angelakis}
\email{dimitris.angelakis@gmail.org}
\affiliation{Science Department, Technical University of Crete, Chania, Crete, Greece, 73100}
\affiliation{Centre for Quantum Technologies, National University of Singapore, 3 Science Drive 2, Singapore 117543}

\author{MingXia Huo}
\affiliation{Centre for Quantum Technologies, National University of Singapore, 3 Science Drive 2, Singapore 117543}

\author{Darrick Chang}
\affiliation{ICFO, Institute of Photonic Sciences, Barcelona, Spain 08860}

\author{Leong Chuan Kwek}
\affiliation{Centre for Quantum Technologies, National University of Singapore, 3 Science Drive 2, Singapore 117543}
\affiliation{Institute of Advanced Studies (IAS) and National Institute of Education,
Nanyang Technological University, Singapore 639673}

\author{Vladimir Korepin}
\affiliation{C. N. Yang Institute for Theoretical Physics, State University of New York at
Stony Brook, NY 11794-3840, USA}

\begin{abstract}
One of the most well known relativistic field theory models is the Thirring
model (TM). Its realization can demonstrate the famous prediction for the
renormalization of mass due to interactions.
However, experimental verification of the latter requires complex  
accelerator experiments whereas  analytical solutions of the model
can be extremely cumbersome to obtain.
 In this work, following Feynman's original proposal,
 we propose a alternative quantum system as a simulator
 of the TM dynamics. Here the relativistic particles are mimicked,
counter-intuitively, by polarized photons in a quantum nonlinear medium.
We show that the entire set of regimes of the Thirring model -- bosonic or fermionic, and massless or massive -- can be faithfully reproduced using coherent light trapping techniques. The sought
after correlations' scalings can be extracted by simple probing of the coherence functions
of the light using standard optical techniques.

\end{abstract}

\maketitle

\subsection*{Introduction}

The Thirring model is the simplest relativistic quantum field theory that
can describe the self interaction of a Dirac field in 1+1 dimensions \cite{Thirring}. 
 Various techniques have been developed to predict its
correlations function behaviour with conclusive results only for the
massless fermonic case \cite{Coleman,Klaiber}. For the opposite case of
fermions with mass, approximations involving the form factors approach have been
developed for the n-point correlations whereas a full analytical 
solution still evades the efforts of seminal theoretical physicists \cite{Zamo,Smirnov, Fring, Cardy,Korepin}.
The theory of the bosonic regime and the correlation function behavior is  less
developed with efforts mainly concentrated in the solving the classical case\cite{Mihalov,Bhattacharya}.

The possibility to physically simulate the Thirring model dynamics in a different
controllable quantum system is thus extremely interesting. An ideal quantum
simulation should be capable of reproducing on demand the bosonic or the
fermionic cases with tunable interactions and mass terms. In addition direct
access to the correlation function behaviour will be an advantage. Such a
system would not only shed light in the unknown parts of the Thirring model
itself, but motivate works for probing of the behaviour of other
relativistic quantum field theories which are beyond the realm of numerical
and experimental techniques. In this work we describe exactly such a setup
where the TM is generated counter-intuitively in nonlinear optical system.
Here polarized photons are mimicking the behaviour of the relativistic
fermions where interactions and the mass terms are generated and controlled
using quantum slow light techniques.

Recent works in quantum simulations of relativistic theories focused on free
fermion simulations using cold atoms and ion systems 
\cite{LewensteinAhufinger12, BlochDalibard12, BlattRoss12}. An
interacting fields proposal is described in a recent seminal work where
fermions on an optical lattice are used \cite{Cirac} and more recently a  connection
between a variational simulation of fields to a generic cavity QED setup was made\cite{Barrett}. The present work
differs in many aspects to all previous quantum simulations. First and foremost it is not
based on cold atoms or particles with mass in general, but on a fundamentally
different and somewhat counter intuitive approach using photons in a
nonlinear optical medium. In addition, our optical setup being a natural
continuum system, has the ability to reproduce the continuous Thirring model
directly without resorting to a lattice approach. Finally, both the bosonic
and the fermionic cases can be generated by controlling the relevant optical
interactions, and in both massless and massive regimes. In our proposal the
relativistic dynamics (linear dispersion) and the necessary strong
interactions are generated and controlled by employing electromagnetically
induced transparency (EIT) techniques \cite{EIT,DPS1,DPS2}.The 
scalings of the correlation functions of the TM, for any regime of
interactions, could be simply probed by analyzing the quantum optical
coherence functions of the outgoing photons as they exit the medium.

We start our analysis by showing the possibility to generate a nonlinear
Dirac type of Hamiltonian using stationary polarized pulses of light. We
then analyze the parameter regime to tune the system to a hardcore regime
reproducing the FTM dynamics and continue by showing that both the massless
and massive cases are reproducible. We conclude by discussing how one could
probe the scaling of correlation functions for any regime of interactions
and mass values using standard quantum optical measurements. In this last
part, we calculate analytically for demonstration purposes the correlation
function behaviour for the case of the massless fermionic TM where an exact solution is
known. This could be used to gauge or first test our simulator before it
is used in the territory of theories without known solutions.

\subsection{Photons as interacting Dirac fermions: The system}

\begin{figure}[tbp]
\includegraphics[width=0.98\linewidth]{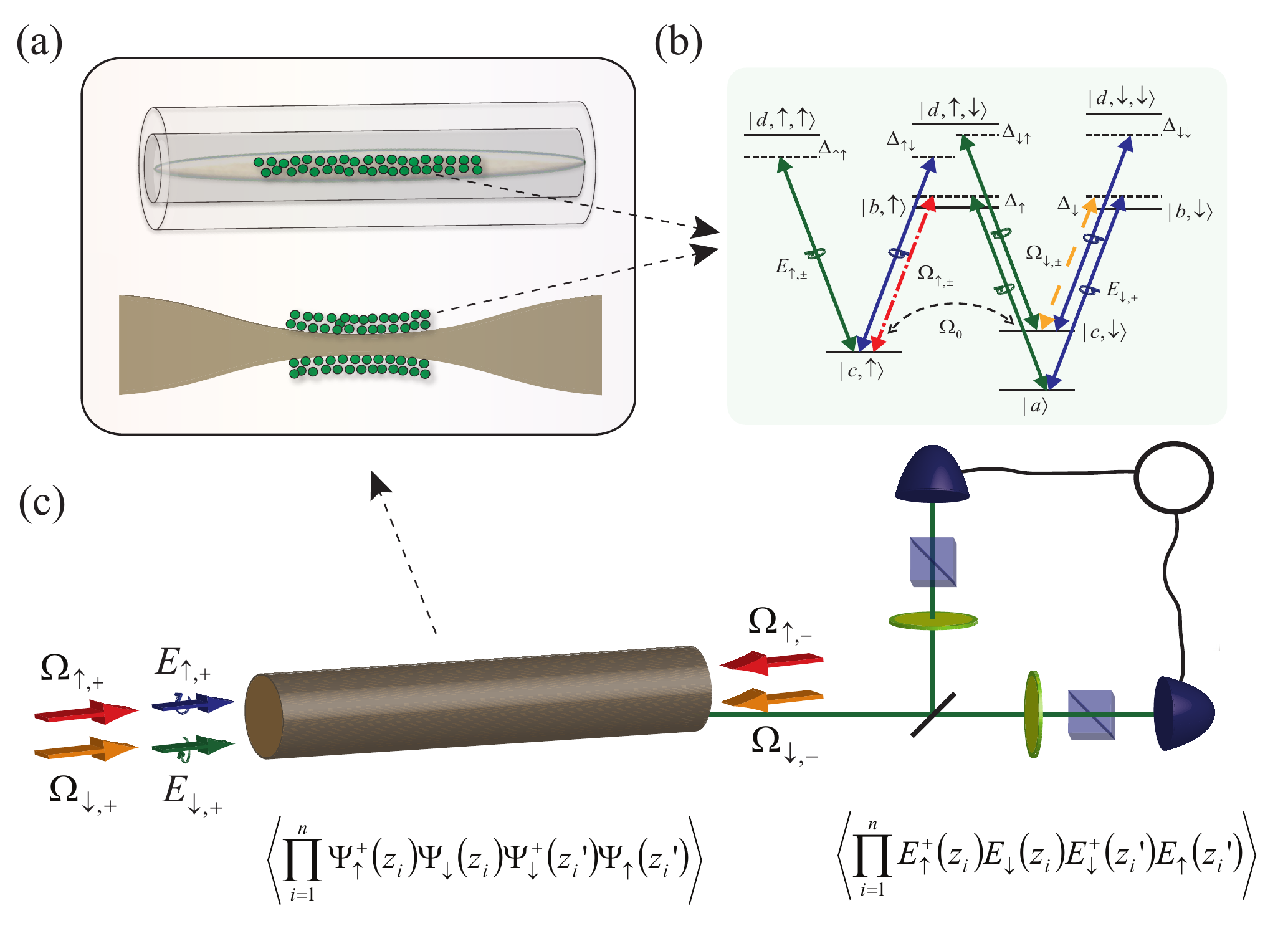}
\caption{An illustration of the possible realization of our proposal.
The quantum pulses $\hat{E}_{\uparrow ,+ }$ and $%
\hat{E}_{\downarrow ,+ }$ enter the  nonlinear waveguide with different
polarizations where they are trapped and made to strongly interact.  
$\Omega _{\uparrow ,\pm }$ and $\Omega _{\downarrow ,\pm }$
are  two pairs of classical lasers helping to trap and control the pulses. 
 In this regime, the polarized pulses can be made
 to mimick interacting fermions of opposite spins 
 following the TM dynamics. The scheme is based in engineering the photons' interaction with atoms in the waveguide using
 EIT techniques. The  appropriate relativistic dispersion relation and the relevant TM mass and  interaction terms, are  fully
 tunable by tuning optical parameters such as atom-photon detunings and laser strengths.
Once the strongly interacting regime is reached, the pulses can be released with all correlations established during the interaction period coherently transfered over to outgoing photon pulses. The latter can be measured using standard quantum optics techniques. At the insets the atomic level structure and a possible implementation using hollow-core fiber or tapered fibers interacting with cold atomic gases.}
\label{fig1}
\end{figure}

Our proposal is based on exploiting the available huge photonic
nonlinearities that can be generated in specific quantum optical setups. More
specifically, we envisage the use of a highly nonlinear waveguide where the
necessary nonlinearity will emerge through the strong interaction of the
propagating photons to existing emitters in the waveguide. Recent efforts
have developed two similar setups in this direction, both capable of
implementing our proposal with either current or near future platforms. In
these experiments, cold atomic ensembles are brought close to the surface of
a tapered fiber \cite{Vetsch10} or are loaded inside the core of a
hollow-core waveguide \cite{Bajcsy11} as shown in Fig. \ref{fig1}\ (a). The
available huge photon nonlinearities due to the (EIT) effect can be used 
to create situations where the trapped photons obey TM dynamics.

We start by assuming two probe quantum fields of few photons each, labeled
as $E_{s}$, $E_{s^{\prime }}$, are tuned to propagate in the medium. $%
s,s^{\prime }=\uparrow ,\downarrow $ will denote their different
polarizations ( later to be representing the spins of the simulated
relativistic particles). The waveguide can also be illuminated by two pairs
of counter-propagating control lasers with Rabi-frequencies $\Omega _{s,\pm }
$ with $\pm \ $ denoting right- and left-moving directions. The atoms placed
inside or close to the surface of the waveguide (depends on the
implementation) have a ground state $\left\vert a\right\rangle $ and excited
states $\left\vert b,s\right\rangle $ and $\left\vert d,s,s^{\prime
}\right\rangle $ Each of the probe pulses is paired with an appropriate
 classical field and is tuned to an atomic
transition such that a typical $\Lambda $ (EIT) configuration is setup
(see Fig. 1(c) and Methods). 

The process to generate the TM dynamics in our photonic system is as
follows: First, the laser-cooled atoms are moved into position so they will
interact strongly with incident quantum light fields as done in \cite{Vetsch10,Bajcsy11}. Initially, resonant to
the corresponding transitions, the classical optical pulses with opposite
polarizations are sent in from one direction, say the left side. They are
injected into the waveguide with the co-propagating classical control fields
$\Omega _{\uparrow ,+}(t)$ and $\Omega _{\downarrow ,+}(t)$ initially turned
on. As soon as the two quantum pulses completely enter into the waveguide,
the classical fields are adiabatically turned off, converting the quantum
probes  into coherent atomic excitations as done routinely in slow-light
experiments. We then adiabatically switch on both $\Omega _{s,+}$\ and $%
\Omega _{s,-}$\ from two sides. The probe pulses become trapped due to the
effective Bragg scattering from the stationary classical waves as analyzed
in \cite{DPS2,Chang08,Angelakis11}. At this stage the pulses are noninteracting with
the photons expanding freely due to the dispersion. By slowly shifting the $d
$-levels we show that the effective masses of the excitations can be kept
constant whereas the effective intra- and interspecies repulsions are
increased. This drives the system into a strongly interacting regime. This
dynamic evolution is possible by keeping for example the corresponding
photon detunings $\Delta _{s}$\ constant while shifting the $d$-level. Once
this correlated state is achieved, 
 the counter-propagating control fields $\Omega_{s,-}$ 
 responsible for trapping the polaritons are slowly turned off. This will release the corresponding quasi-particles by turning them to
propagating photons which will then exit the fiber. As all correlations
established in the previous step are retained, the propagating wave packets
will carry the correlation of the TM dynamics taking place in the medium
outside, where they can be probed by appropriately placed photon detectors.

{\bf A Thirring model of polarized photons:}
In the trapped regime the dark-state polaritons emerge in the form $\Psi
_{s,+}=\frac{\sqrt{n_{z}}g}{\Omega _{s,+}}E_{s,+}$ and $\Psi _{s,-}=\frac{%
\sqrt{n_{z}}g}{\Omega _{s,-}}E_{s,-}$, which are also propagating in both
directions (see Methods). The symmetric combination $\Psi _{s}=\alpha
_{s,+}\Psi _{s,+}+\alpha _{s,-}\Psi _{s,-}$ with $\alpha _{s,\pm }=\frac{%
\Omega _{s,\pm }^{2}}{\Omega _{s,+}^{2}+\Omega _{s,-}^{2}}$ forms a
stationary wave obeying the following tunable nonlinear dynamics (see
Methods)
\begin{eqnarray}
i\hbar \partial _{t}\Psi _{s} &=&-\frac{\hbar ^{2}}{2m_{\mathrm{nr,}s}}%
\partial _{z}^{2}\Psi _{s}+i\hbar \eta _{s}\partial _{z}\Psi _{s}+\hbar
\Omega _{0}\Psi _{\overset{\_}{s}}  \notag \\
&&+\chi _{ss}\Psi _{s}^{\dagger }\Psi _{s}^{2}+\chi _{s\overset{\_}{s}}\Psi
_{\overset{\_}{s}}^{\dagger }\Psi _{\overset{\_}{s}}\Psi _{s}+\mathrm{noise}
\end{eqnarray}%
where the optically tunable inter- and intra- species interactions are given
by
\begin{eqnarray}
\chi _{ss}&&=4\hbar \overline{\Omega }_{s}^{2}/(\Delta _{ss}n_{z}) \\
\chi _{s\overset{\_}{s}}&&=2\hbar \overline{\Omega }_{s}^{2}\left[ 2+\cos
\left( \varphi _{\overset{\_}{s}}-\varphi _{s}\right) \right] /(\Delta _{s%
\overset{\_}{s}}n_{z})
\end{eqnarray}
The mass of the particles in the relativistic regime is $m_{0,s}=-\hbar
\Omega _{0}/\eta _{s}^{2}$and $\eta _{s}=-2v_{s}\cos 2\varphi _{s}$.
We note here that  the above mass is different from the
``dressed" mass due to the interactions.

The quadratic dispersion term corresponding to the ``classical" kinetic
energy can be ignored in the regime of interactions we operate (Methods). In
this case, if we define a spinor field $\mathbf{\Psi }=(\Psi _{\uparrow
},\Psi _{\downarrow })^{T}$, the Hamiltonian generating the last equation
becomes the bosonic Thirring model

\begin{eqnarray}
H &=&\int dz[-i\hbar \left\vert \eta \right\vert \gamma _{1}\partial
_{z}+m_{0}\eta ^{2})\mathbf{\Psi }  \notag \\
&&+\sum_{s}\frac{\chi _{ss}}{2}\Psi _{s}^{\dagger }\Psi _{s}^{\dagger }\Psi
_{s}\Psi _{s}+\frac{\chi }{2}\overset{\_}{\mathbf{\Psi }}\gamma ^{\mu }%
\mathbf{\Psi }\overset{\_}{\mathbf{\Psi }}\gamma _{\mu }\mathbf{\Psi }],
\end{eqnarray}%
where $\overset{\_}{\mathbf{\Psi }}\mathbf{=\Psi }^{\dagger }\gamma _{0}$
and the spinor fields satisfy the commutation relations $[\Psi _{s}\left(
z\right) ,\Psi _{s^{\prime }}\left( z^{\prime }\right) ]=0$, $[\Psi
_{s}\left( z\right) ,\Psi _{s^{\prime }}^{\dagger }\left( z^{\prime }\right)
]=\delta (z-z^{\prime })\delta _{ss^{\prime }}$. The gamma matrices $\gamma
_{\mu }$ for $\mu =0,1$\ are chosen as $\gamma _{0}=\sigma _{x}$, $\gamma
_{1}=i\sigma _{y}$ with covariant gamma matrices $\gamma ^{0}=\gamma _{0}$, $%
\gamma ^{1}=-\gamma _{1}$, and an additional matrix $\gamma _{5}=\gamma
_{0}\gamma _{1}$.

{\bf A fermionic TM with tunable mass:} The Hamiltonian in Eq.(4)
describes bosonic particles made of stationary light-matter excitations.
 In order to realize the FTM we will tune to the
hardcore regime where the same species interaction is much larger than all
other interaction terms. In Fig. 2a we calculate and plot the ratio of their
strength as a function of the controllable single photon detunings. We see
that by appropriate tuning from the corresponding upper atomic states (
tuning the laser towards or away from the corresponding states) the system
enters a hardcore regime where the same species interaction is much larger
than any other term in the Hamiltonian. The particles will then behave as
effective fermions in all aspects-identical density density correlations\cite{Girardeau,Tonks}  and
spectra but different first order correlations due to the sign issue. We note here that as we will discuss later,  our optical simulator will be naturally probing intensity-intensity correlations which nicely fits with the above point.

 For this regime, we also calculate and plot the
individual interaction strengths, denoted  $\protect\beta _{ss}^{\mathrm{i}}$
$(\protect\beta _{s\protect\overset{\_}{s}}^{\mathrm{i}})$, as a function of the kinetic energy of the
particles for the same values of the tunable optical detunings (Fig. 2b and
2c). We see that when the system is optically tuned to the hardcore regime,
the intra-species repulsion is much larger not only the inter-species
interaction (Fig2 a) but the kinetic energy as well (Fig2b,c). In plotting (Fig2 a)
we have set $m_{0}=0$ with $\Omega _{0}$ turned off. We note here the simple
way of tuning the rest mass of the particles by simply controlling the
\textquotedblleft connecting" laser. 

\begin{figure}[tbp]
\includegraphics[width=0.99\linewidth]{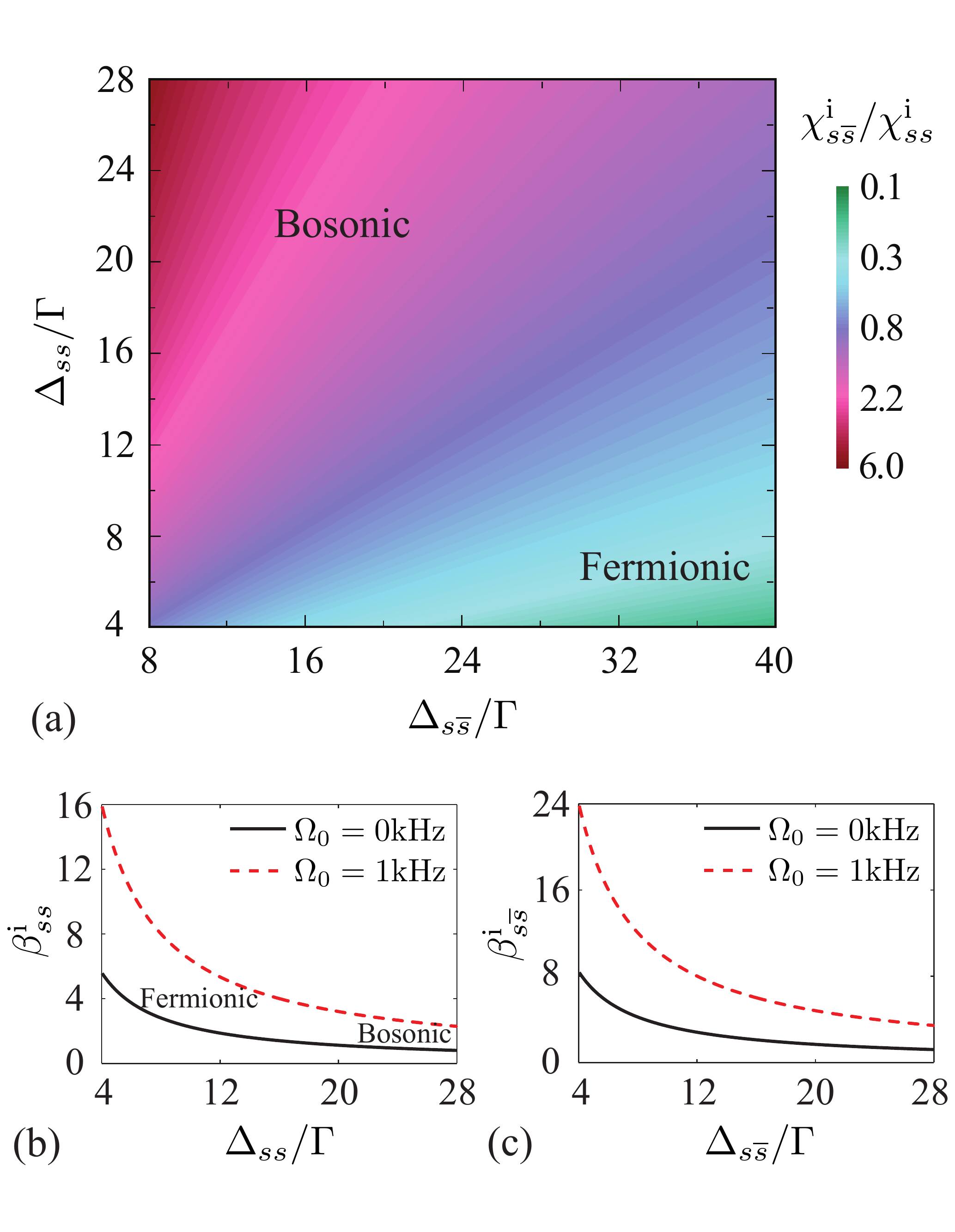} %
\caption{Fig 2a: The ratio of inter- to intra species repulsion as a
function of the relevant single photon detunings and the relevant "bosonic"
and "fermionic" regimes; Fig 2b(c): The ratio of intra(inter) species
interaction to kinetic energy, denoted by $\protect\beta _{ss}^{\mathrm{i}}$
$(\protect\beta _{s\protect\overset{\_}{s}}^{\mathrm{i}})$, as a function of
the corresponding single photon detuning }
\label{fig2}
\end{figure}

\subsection{Probing the TM correlation scalings with photon detectors}

In the following we describe how one could probe the scaling behaviour of
the correlation functions and from that infer the model's properties and the
fascinating renormalization of mass due to the interactions. We use the case
of the massless fermionic TM as an example to illustrate  our approach. Of
course in this case the spectrum of the model and the scattering matrix was
explicitly evaluated by the Bethe Ansatz \cite{Korepin} and is known. 
Nevertheless, this regime serves as an excellent check for our quantum simulator. 
 Ultimately, one would like to move to the relatively unexplored regimes of massive or bosonic TM as well as other interesting relativistic theories\footnote{ We note
here that to our knowledge although the integrability of BTM has been
investigated in several works connected with the NSE and
the sine-Gordon (sG) model \cite{Bhattacharya}, evaluating its correlation
functions is hardly resolved.}

 In the massless FTM  
 the long-scaled n-point correlation as a function of the distance $z-z^{\prime }$ and
interaction strength $\chi /\left\vert \eta \right\vert $\ was calculated by
Coleman in \cite{Coleman} and reads:
\begin{eqnarray}
&&\left\langle 0\right\vert \prod_{i=1}^{n}\Psi _{\uparrow }^{\dagger
}\left( z_{i}\right) \Psi _{\downarrow }\left( z_{i}\right) \Psi
_{\downarrow }^{\dagger }\left( z_{i}^{\prime }\right) \Psi _{\uparrow
}\left( z_{i}^{\prime }\right) \left\vert 0\right\rangle   \notag \\
&=&\left( \frac{1}{2}\right) ^{2n}\frac{\Lambda ^{2}\prod_{i>j}\left[ \left(
z_{i}-z_{j}\right) ^{2}\left( z_{i}^{\prime }-z_{j}^{\prime }\right)
^{2}M^{4}\right] ^{[1+\frac{\chi }{\pi \left\vert \eta \right\vert }]^{-1}}}{%
\prod_{i,j}\left[ \Lambda ^{2}\left( z_{i}-z_{j}^{\prime }\right) ^{2}\right]
^{[1+\frac{\chi }{\pi \left\vert \eta \right\vert }]^{-1}}},  \notag \\
&&
\end{eqnarray}%
One can extract the two-point correlation with $(n=1)$:%
\begin{eqnarray}
&&\left\langle 0\right\vert \Psi _{\uparrow }^{\dagger }\left( z_{1}\right)
\Psi _{\downarrow }\left( z_{1}\right) \Psi _{\downarrow }^{\dagger }\left(
z_{1}^{\prime }\right) \Psi _{\uparrow }\left( z_{1}^{\prime }\right)
\left\vert 0\right\rangle   \notag \\
&=&\frac{\Lambda ^{2}}{4}\left[ \Lambda ^{2}\left( z_{1}-z_{1}^{\prime
}\right) ^{2}\right] ^{-[1+\frac{\chi }{\pi \left\vert \eta \right\vert }%
]^{-1}}.  \label{TM_cor}
\end{eqnarray}%
$\Lambda $ is the momentum cutoff, which we can estimate
for our finite size waveguide of lentgh L containing N interacting
particles, to be $\Lambda =\pi n_{\mathrm{ph}}\sin \left( \chi /\left\vert
\eta \right\vert \right) /\left( \chi /\left\vert \eta \right\vert \right) $
, where $n_{ph}=N/L$ \cite{Korepin}. In Fig. 3(a) we plot the dependance
of the cutoff on the ratio of interactions to kinetic energy for the regime $%
0<\chi /\left\vert \eta \right\vert <\pi $ where modes with unphysically
high energy can be excluded. This regime of interactions corresponds to our optical simulator in
 single photon detuning $\Delta _{ss^{\prime }}$ ranging up to a few $\Gamma $(which is
within the required conditions for our slow light setup dynamics \cite
{EIT,DPS1}). In Fig. 3(b), we numerically calculate and plot the two-point correlation 
as expressed in Eq. (\ref{TM_cor}) as a function of the distance
normalized to the inverse of the photonic density for maximum interactions.
In these plots, as also in the previous one in Fig.2, we have assumed the
typical values of the optical parameters for the slow light setup, ie. $%
\left\vert \cos 2\varphi _{s}\right\vert =0.004$, $n_{z}=10^{7}\mathrm{m}^{%
\mathrm{-1}}$, $\Gamma _{\mathrm{1D}}=0.2\Gamma $, and $\overline{\Omega }%
_{s}\simeq 1.5\Gamma $, $0<\chi /\left\vert \eta \right\vert <\pi $\

\begin{figure}[tbp]
\includegraphics[width=0.48\linewidth]{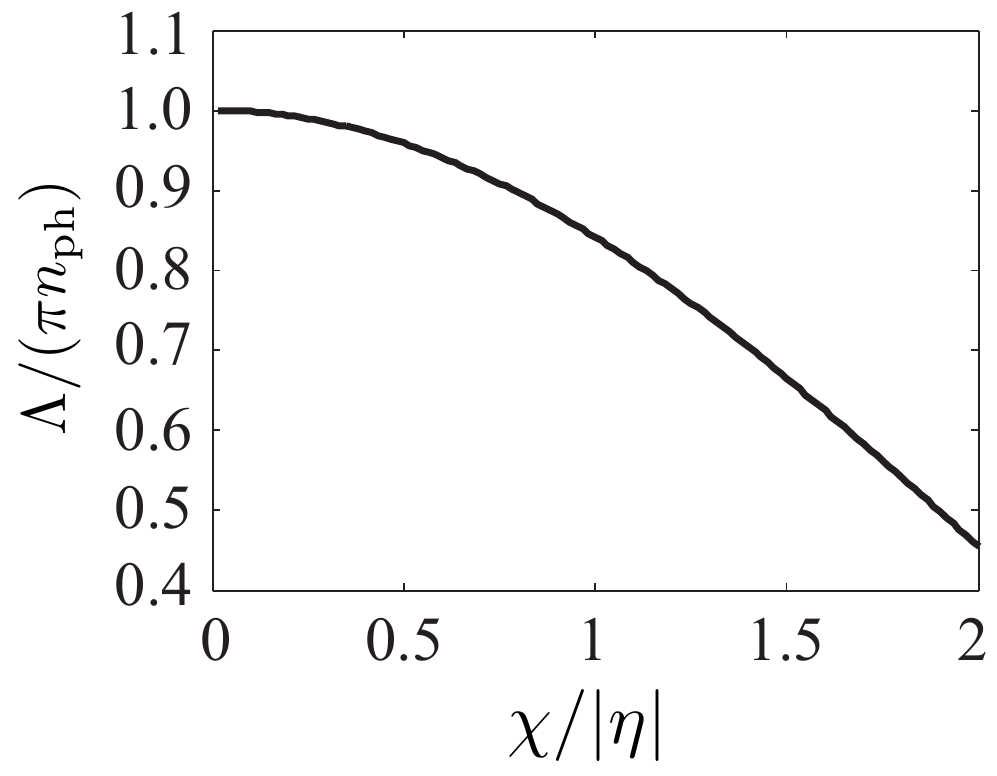} %
\includegraphics[width=0.49\linewidth]{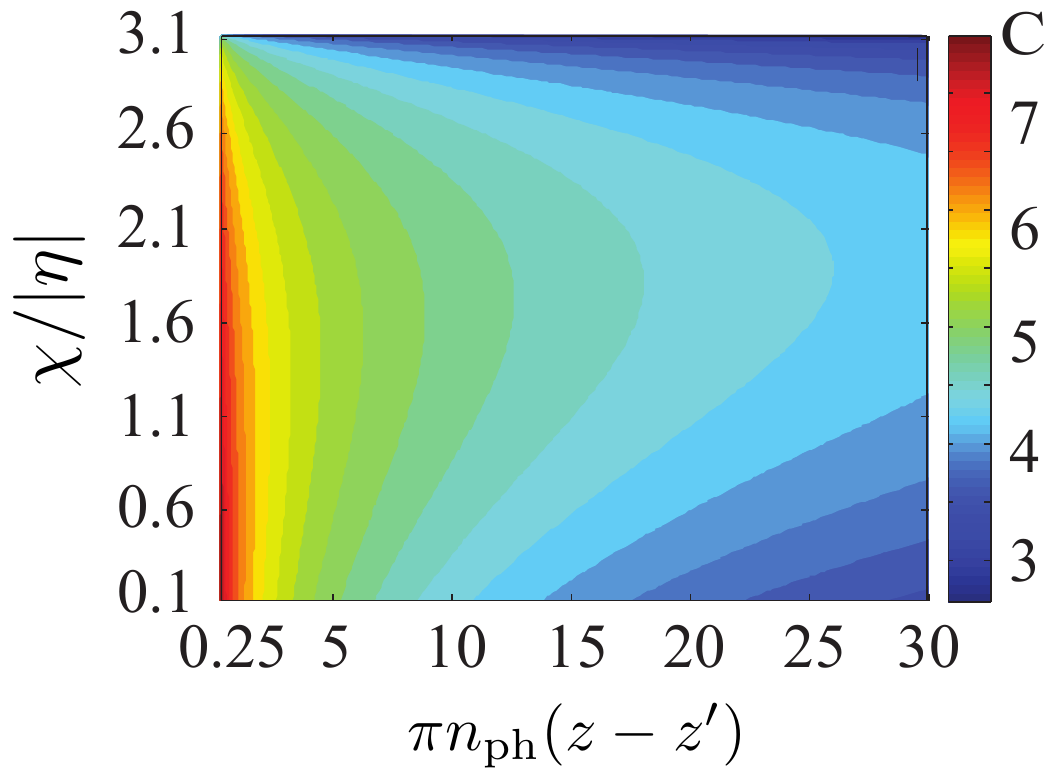}
\caption{The momentum cutoff $\Lambda $\ in unit of $\protect\pi n_{\mathrm{%
ph}}$\ (a) and log-scaled two-point correlations (b) of the massive FTM. As
the distances $z-z^{\prime }$ in (b) increases, the correlations decrease.
We simulate the system that each initial quantum pulse contains $10$ photons
and spreads over after completely entering into $1\mathrm{cm}$-length\
fiber. }
\label{fig3}
\end{figure}

To detect the above correlation in our optical proposal, is enough to observe
that it is in fact an effective transverse spin-spin correlation
function $\left\langle S^{+}\left( z_{1}\right) S^{-}\left( z_{1}^{\prime
}\right) \right\rangle $ with $S^{+}\left( z_{1}\right) =\Psi _{\uparrow
}^{\dagger }\left( z_{1}\right) \Psi _{\downarrow }\left( z_{1}\right) $ and
$S^{-}\left( z_{1}^{\prime }\right) =\Psi _{\downarrow }^{\dagger }\left(
z_{1}^{\prime }\right) \Psi _{\uparrow }\left( z_{1}^{\prime }\right) $. 
In our case, one would release and map the polaritons to photons by turning off $\Omega _{s,-}
$, and then using polarization wave plates to convert $\Psi _{s}$ into their
superpositions $\Psi _{x,\pm }$ and $\Psi _{y,\pm }$\  as the pulses exit
the medium which then can be measured by probing intensity-intensity correlations(see Methods).
The latter  could be done using standard
Handbury-Brown and Twiss type of techniques for different times as
 time correlations for the propagating photons here map to space of the interacting particles.

{\bf Conclusions:} 
We have shown how to physically simulate the Thirring model dynamics in a photonic setup where
fermions are mimicked by stationary polarized pulses of light in a strongly nonlinear EIT medium. 
We showed in detail how using coherent light trapping techniques, 
the photons' dispersion relation can be tuned to the relativistic regime
and how to create optically tunable interactions and mass terms. The capability to naturally probe
the scaling behaviour of the TM correlation functions
 using standard optical measurements on the emitted pulses, makes our approach complementary
 to alternative simulations based in cold atom setups.
The possibility to first test the device in regimes
 where analytical solutions are known (massless regime) and then probe
 regimes with unknown solution (massive FTM and bosonic TM) or other interacting relativistic QFTs 
 makes this extremely interesting in our opinion. Finally, the open nature of our setup and the interaction
with the environment through photon loss and driving,
 should allow for the possibility to investigate QFT theories at finite temperature,
 an extremely fascinating and well unexplored area of research. We hope that the above
characteristics combined with the simplicity of our approach will inspire further experimental and theoretical
 investigations in using photonic systems for quantum simulations of exotic theories.

{\bf Acknowledgements:} We would like to acknowledge financial support by the National Research Foundation \& Ministry of Education, Singapore. DEC acknowledges support from Fundacio Privada Cellex Barcelona and VK  from NSF DMS 1205422


\section{Methods}

\subsection{Nonlinear dynamics of relativistic stationary polaritons}

The Hamiltonian of the system reads $H=-\hbar \int n_{z}dz(h_{\Delta }+h_{%
\mathrm{Q}}+h_{\mathrm{C}})$, where
\begin{eqnarray}
h_{\Delta } &=&\sum_{s}\Delta _{s}\sigma _{b,s;b,s}+\sum_{s,s^{\prime
}}\Delta _{ss^{\prime }}\sigma _{d,s,s^{\prime };d,s,s^{\prime }}, \\
h_{\mathrm{Q}} &=&\sum_{s}(g_{s}\sigma _{b,s;a}+\sum_{s^{\prime
}}g_{ss^{\prime }}\sigma _{d,s,s^{\prime };c,s})E_{s}+\mathrm{h.c.}, \\
h_{\mathrm{C}} &=&\sum_{s}\sigma _{c,s;b,s}\Omega _{s}+\Omega _{0}\sigma
_{c,\uparrow ;c,\downarrow }+\mathrm{h.c.}.
\end{eqnarray}%
Here, the quantum filed is composed of two counter-propagating components, $%
E_{s}=E_{s,+}(z,t)e^{ik_{\mathrm{Q},s}z}+E_{s,-}(z,t)e^{-ik_{\mathrm{Q},s}z}$%
.\ Similarly, control fields $\Omega _{s}=\Omega _{s,+}(z,t)e^{ik_{\mathrm{C}%
,s}z}+\Omega _{s,-}(z,t)e^{-ik_{\mathrm{C},s}z}$, where $\Omega _{s,\pm }$\
are slowly varying parameters of $z$ and $t$. $k_{\mathrm{Q},s}$ and $k_{%
\mathrm{C},s}$ denote the wave vectors corresponding to central frequencies $%
\omega _{\mathrm{Q},s}$ and $\omega _{\mathrm{C},s}$ of $E_{s,\pm }$ and $%
\Omega _{s,\pm }$, respectively. The collective and continuous operators $%
\sigma _{\mu ;\nu }\equiv \sigma _{\mu ;\nu }(z,t)$ describe the averages of
the flip operators $\left\vert \mu \right\rangle \left\langle \nu
\right\vert $ over atoms in a small region around $z$. $n_{z}$ is the atomic
density. $g_{s}$ and $g_{ss^{\prime }}$ are coupling strengths between the
quantum fields and atoms, while $\Delta _{s}$ and $\Delta _{ss^{\prime }}$
are one-photon detunings for corresponding transitions. For simplicity, we
assume that $g_{s}=g_{ss^{\prime }}=g$.

As done usually in slow light works\cite{DPS1,DPS2}, we consider that atoms are
initialized in the ground state $\left\vert a\right\rangle $. In the trapped
regime the dark-state polaritons emerge in the form $\Psi _{s,+}=\frac{\sqrt{%
n_{z}}g}{\Omega _{s,+}}E_{s,+}$ and $\Psi _{s,-}=\frac{\sqrt{n_{z}}g}{\Omega
_{s,-}}E_{s,-}$, which are also propagating in both directions. Here, we
have considered the slow-light limit where the control fields are tuned to
be very weak and the polaritons are mostly spin excitations. The imbalance
between two counter-propogating components of control fields is described by
$\tan ^{2}\varphi _{s}=\Omega _{s,-}^{2}/\Omega _{s,+}^{2}$. Another angular
parameter is introduced as $\tan ^{2}\theta _{s}=g^{2}n_{z}/\overline{\Omega
}_{s}^{2}$ with $\overline{\Omega }_{s}^{2}=(\Omega _{s,+}^{2}+\Omega
_{s,-}^{2})/2$. The reduced light velocity in the nonlinear medium\ is $%
v_{s}=v/(\pi \tan ^{2}\theta _{s})$, where $v=\omega _{\mathrm{Q},s}/k_{%
\mathrm{Q},s}$ is the light speed in the empty medium. The dark-state
polaritons dynamics is determined by the Maxwell-Bloch equations of quantum
field
\begin{eqnarray}
(\partial _{t}\pm v\partial _{z})E_{s,\pm }=in_{z}g(\sigma _{a;b,s;\pm
}+\sum_{s^{\prime }}\sigma _{c,s;d,s,s^{\prime };\pm })
\end{eqnarray}

Substituting the dark-state operators $\Psi _{s,\pm }$ and adiabatically
eliminating the fast decaying atomic operators and ignoring high-frequency
oscillation terms, the Maxwell-Bloch equations become

\begin{eqnarray}
&&\partial _{t}\Psi _{s}+v\partial _{z}(\alpha _{s,+}\Psi _{s}-\alpha
_{s,-}\Psi _{s}+2\alpha _{s,+}\alpha _{s,-}A_{s})  \notag \\
&\simeq &-\frac{\tan ^{2}\theta _{s}}{2}\partial _{t}\Psi _{s}+i\frac{\tan
^{2}\theta _{s}}{2}\Omega _{0}\Psi _{\overset{\_}{s}}-i\frac{2g^{2}}{\Delta
_{ss}}\Psi _{s}^{\dagger }\Psi _{s}\Psi _{s}  \notag \\
&&-i\frac{g^{2}}{\Delta _{s\overset{\_}{s}}}\left[ 2+\cos \left( \varphi _{%
\overset{\_}{s}}-\varphi _{s}\right) \right] \Psi _{\overset{\_}{s}%
}^{\dagger }\Psi _{\overset{\_}{s}}\Psi _{s},  \notag \\
&&\partial _{t}A_{s}+v\partial _{z}(2\Psi _{s}+\alpha _{s,-}A_{s}-\alpha
_{s,+}A_{s})  \notag \\
&\simeq &-i\frac{n_{z}g^{2}}{\Delta _{s}}A_{s}-\frac{ig^{2}}{\Delta _{ss}}%
\Psi _{s}^{\dagger }\Psi _{s}A_{s}-\frac{ig^{2}}{\Delta _{s\overset{\_}{s}}}%
\Psi _{\overset{\_}{s}}^{\dagger }\Psi _{\overset{\_}{s}}A_{s}  \notag \\
&&-\frac{ig^{2}\Omega _{\overset{\_}{s},+}}{\Delta _{s\overset{\_}{s}}\Omega
_{s,+}}\Psi _{\overset{\_}{s}}^{\dagger }\Psi _{s}A_{\overset{\_}{s}}.
\end{eqnarray}

Here $s,\overset{\_}{s}=\uparrow ,\downarrow $ and $s\neq \overset{\_}{s}$.
We keep the symmetric combination $\Psi _{s}=\alpha _{s,+}\Psi _{s,+}+\alpha
_{s,-}\Psi _{s,-}$ with $\alpha _{s,\pm }=\frac{\Omega _{s,\pm }^{2}}{\Omega
_{s,+}^{2}+\Omega _{s,-}^{2}}$\ while the antisymmetric combination $A_{s}$
proportional to $\Psi _{s,+}-\Psi _{s,-}$ can be adiabatically eliminated in
the limit of large optical depth since the pulse matching phenomenon gives $%
\Psi _{s,+}-\Psi _{s,-}\rightarrow 0$ \cite{DPS2}.

The evolution equations for dark-state polaritons then become
\begin{eqnarray}
i\hbar \partial _{t}\Psi _{s} &=&-\frac{\hbar ^{2}}{2m_{\mathrm{nr,}s}}%
\partial _{z}^{2}\Psi _{s}+i\hbar \eta _{s}\partial _{z}\Psi _{s}+\hbar
\Omega _{0}\Psi _{\overset{\_}{s}}  \notag \\
&&+\chi _{ss}\Psi _{s}^{\dagger }\Psi _{s}^{2}+\chi _{s\overset{\_}{s}}\Psi
_{\overset{\_}{s}}^{\dagger }\Psi _{\overset{\_}{s}}\Psi _{s}+\mathrm{noise}
\label{NLE}
\end{eqnarray}%
with interspecies interaction strength given by $\chi _{ss}=4\hbar \overline{%
\Omega }_{s}^{2}/(\Delta _{ss}n_{z})$, while the intraspecies interaction
strength is $\chi _{s\overset{\_}{s}}=2\hbar \overline{\Omega }_{s}^{2}\left[
2+\cos \left( \varphi _{\overset{\_}{s}}-\varphi _{s}\right) \right]
/(\Delta _{s\overset{\_}{s}}n_{z})$. $m_{\mathrm{nr,}s}=-\hbar \overline{%
\Omega }_{s}^{2}/[4\sin ^{2}(2\varphi _{s})v_{s}^{2}\Delta _{s}]$ is the
polariton mass. The mass of the particles in the relativistic regime is $%
m_{0,s}=-\hbar \Omega _{0}/\eta _{s}^{2}$, which is different from the
non-relativistic polariton mass $m_{\mathrm{nr,}s}$, and $\eta
_{s}=-2v_{s}\cos 2\varphi _{s}$. Note that  the above is different from the
``dressed" mass due to the interactions.

\subsection*{The bosonic Thirring model}

We will now show that we can tune our system such that the first term on the
r.h.s. of the equation, the quadratic kinetic energy term, to be much smaller
 than the linear relativistic one (second term). For this
we first set  $\eta _{\uparrow }=-\eta_{\downarrow }=\eta $ to simplify the
expressions. The latter is possible by tuning  $\theta _{\uparrow}=\theta
_{\downarrow }$ through the slow light velocities for each species $%
v_{\uparrow }=v_{\downarrow }$\ and $\cos 2 \varphi _{\uparrow }=-\cos
2\varphi _{\downarrow }$. These correspond to different ratio Rabi
frequencies of the classical lasers $\Omega _{s,+}^{2}$ and $\Omega
_{s,-}^{2}$ set by $\tan ^{2}\varphi _{s}$. The non-relativistic kinetic
energy reads
\begin{equation}
E_{\mathrm{nr,}s}=\left( P_{s}\right) ^{2}/(2m_{\mathrm{nr,}s})=2\hbar \sin
^{2}(2\varphi _{s})v_{s}^{2}\left\vert \Delta _{s}\right\vert /[\overline{%
\Omega }_{s}^{2}\left( z_{s}\right) ^{2}]
\end{equation}
and corresponds to the first quadratic dispersion relation term above. The
relativistic one (second term in r.h.s. of Eq. 6) is, $E_{r,s}=|\eta_s| P_s$, where $P_s$ is the characteristic momentum of the polaritons. One can estimate $P_s \sim \hbar/z_s$ based on the typical spatial extent of the polaritons $z_s$, which can range between $L$ for linear or weakly interacting photons to $L/N_{ph,s}$ in the Tonks gas regime. The latter
ranges from L to $L/N_{\mathrm{ph,}s}$ ( weak or linear to Tonks gas regime)
with $N_{\mathrm{ph,}s}$ is the $s$-type photon number. Setting their ratio
to $\beta _{s}^{\mathrm{k}}$, we can easily see that the latter is simple
tunable function of one-photon detuning:
\begin{equation}
\beta _{s}^{\mathrm{k}}=\frac{E_{\mathrm{nr,}s}}{E_{\mathrm{r,}s}}=\frac{%
\sin ^{2}(2\varphi _{s})v_{s}\left\vert \Delta _{s}\right\vert }{\left\vert
\cos 2\varphi _{s}\right\vert z_{s}\overline{\Omega }_{s}^{2}},
\end{equation}%
Fixing the mixing angle $\left\vert \cos 2\varphi _{s}\right\vert =0.004$,
the atomic and photonic densities $n_{\mathrm{ph,}s}=n_{\mathrm{ph}}=10^{3}%
\mathrm{m}^{\mathrm{-1}}$, the cooperativity factor $\Gamma _{\mathrm{1D}%
}=0.2\Gamma $, $v_{s}=100\mathrm{m/s}$\ and $\overline{\Omega }_{s}\simeq
1.5\Gamma $, $\beta _{\mathrm{k,}s}$;, we see that the ratio can be as small
as $5\%$ for similar values of photon detuning $\Delta _{s}/\Gamma $ from $%
0.05$ to $0.08$, which allows us to neglect the contribution of $E_{\mathrm{%
nr,}s}$ in the Hamiltonian Eq. (\ref{model}). The nonlinear equation then
becomes the bosonic TM:
\begin{eqnarray}
H &=&\int dz[-i\hbar \left\vert \eta \right\vert \gamma _{1}\partial
_{z}+m_{0}\eta ^{2})\mathbf{\Psi }  \notag \\
&&+\sum_{s}\frac{\chi _{ss}}{2}\Psi _{s}^{\dagger }\Psi _{s}^{\dagger }\Psi
_{s}\Psi _{s}+\frac{\chi }{2}\overset{\_}{\mathbf{\Psi }}\gamma ^{\mu }%
\mathbf{\Psi }\overset{\_}{\mathbf{\Psi }}\gamma _{\mu }\mathbf{\Psi }],
\label{model}
\end{eqnarray}%
where $\overset{\_}{\mathbf{\Psi }}\mathbf{=\Psi }^{\dagger }\gamma _{0}$
and $\mathbf{\Psi }=(\Psi _{\uparrow },\Psi _{\downarrow })^{T}$ with the
spinor fields satisfying the commutation relations $[\Psi _{s}\left(
z\right) ,\Psi _{s^{\prime }}\left( z^{\prime }\right) ]=0$, $[\Psi
_{s}\left( z\right) ,\Psi _{s^{\prime }}^{\dagger }\left( z^{\prime }\right)
]=\delta (z-z^{\prime })\delta _{ss^{\prime }}$. The gamma matrices $\gamma
_{\mu }$ for $\mu =0,1$\ are chosen as $\gamma _{0}=\sigma _{x}$, $\gamma
_{1}=i\sigma _{y}$ with covariant gamma matrices $\gamma ^{0}=\gamma _{0}$, $%
\gamma ^{1}=-\gamma _{1}$, and an additional matrix $\gamma _{5}=\gamma
_{0}\gamma _{1}$. We have also set$\ m_{0}=m_{0,\uparrow }=m_{0,\downarrow }$%
, which is achievable by tuning $\eta _{\uparrow }=-\eta _{\downarrow }$.\
Furthermore  we can also have $m_{\mathrm{nr,}\uparrow }=m_{\mathrm{nr,}%
\downarrow }=m_{\mathrm{nr}}$\ and $2\chi _{\uparrow \downarrow }=2\chi
_{\uparrow \downarrow }=\chi $ by appropriately tuning the single and two
photon detunings.



\subsection{Probing of the TM correlations optically}


The two-point correlation
function of massless FTM is an effective transverse spin-spin correlation
function $\left\langle S^{+}\left( z_{1}\right) S^{-}\left( z_{1}^{\prime
}\right) \right\rangle $ with $S^{+}\left( z_{1}\right) =\Psi _{\uparrow
}^{\dagger }\left( z_{1}\right) \Psi _{\downarrow }\left( z_{1}\right) $ and
$S^{-}\left( z_{1}^{\prime }\right) =\Psi _{\downarrow }^{\dagger }\left(
z_{1}^{\prime }\right) \Psi _{\uparrow }\left( z_{1}^{\prime }\right) $. We
could probe this by measuring correlations between the two polaritons
species densities $\rho _{x,\pm }=\Psi _{x,\pm }^{\dagger }\Psi _{x,\pm }$
and $\rho _{y,\pm }=\Psi _{y,\pm }^{\dagger }\Psi _{y,\pm }$ with $\Psi
_{x,\pm }=(\Psi _{\uparrow }\pm \Psi _{\downarrow })/\sqrt{2}$ and $\Psi
_{y,\pm }=(\Psi _{\uparrow }\mp i\Psi _{\downarrow })/\sqrt{2}$. We can
express $S^{+}=\Psi _{\uparrow }^{\dagger }\Psi _{\downarrow }=[(\rho
_{x,+}-\rho _{x,-})+i(\rho _{y,+}-\rho _{y,-})]/2$ and $S^{-}=\Psi
_{\downarrow }^{\dagger }\Psi _{\uparrow }=[(\rho _{x,+}-\rho _{x,-})-i(\rho
_{y,+}-\rho _{y,-})]/2$, which indicates the correlation function $%
\left\langle S^{+}\left( z_{1}\right) S^{-}\left( z_{1}^{\prime }\right)
\right\rangle $ of massless FTM as a combination of density-density
correlations $\left\langle \rho _{\alpha ,\pm }\left( z_{1}\right) \rho
_{\alpha ^{\prime },\pm }\left( z_{1}^{\prime }\right) \right\rangle $ with $%
\alpha $, $\alpha ^{\prime }=x$, $y$. 

\subsection*{Losses}

 The losses mainly due to the
spontaneous emission from the atomic upper levels can be estimated by adding
the imaginary parts as $\overrightarrow{\chi }_{ss}=8\hbar \overline{\Omega }%
_{s}^{2}/[n_{z}(2\left\vert \Delta _{ss}\right\vert +i\Gamma )]$ and $%
\overrightarrow{\chi }_{s\overset{\_}{s}}=4\hbar \left[ 2+\cos \left(
\varphi _{\overset{\_}{s}}-\varphi _{s}\right) \right] \overline{\Omega }%
_{s}^{2}/[n_{z}(2\Delta _{s\overset{\_}{s}}+i\Gamma )]$, which leads to the
total loss rate $\kappa =\sum_{s}(\kappa _{ss}+\kappa _{s\overset{\_}{s}})$
with $\kappa _{ss}=8n_{\mathrm{ph}\text{\textrm{,}}s}\overline{\Omega }%
_{s}^{2}\Gamma /\left[ n_{z}\left( 4\Delta _{ss}^{2}+\Gamma ^{2}\right) %
\right] $ and $\kappa _{s\overset{\_}{s}}=4\left( 2+\cos \left( \varphi _{%
\overset{\_}{s}}-\varphi _{s}\right) \right) n_{\mathrm{ph,}s}\overline{%
\Omega }_{s}^{2}\Gamma /\left[ n_{z}\left( 4\Delta _{s\overline{s}%
}^{2}+\Gamma ^{2}\right) \right] $. With the smallest detunings $\Delta
_{ss^{\prime }}\simeq 4\Gamma $ used in the plotting of Fig. 2 for example,
the shortest coherence time $1/\kappa $ is $400\mathrm{\mu s}$. For larger
detunings, the coherence time will be longer. %


\begin{thebibliography}{99}

\bibitem{Thirring} W. Thirring, Annals Phys. \textbf{3}, 91 (1958).

\bibitem{Coleman} S. Coleman, Phys. Rev. D \textbf{11}, 2088-2097 (1975).


\bibitem{Klaiber} B. Klaiber, \textit{Lectures in theoretical physics},
edited by A. Barut and W. Brittin, Vol. X (Gordon and Breach, New York,
1968) part A.


\bibitem{Korepin} V.E. Korepin, N.M. Bogoliubov, and A.G. Izergin, \textit{%
Quantum Inverse Scattering Method and Correlation Functions} (Cambridge
University 1993).

\bibitem{Zamo}
V.Fateev, et al., Nucl.Phys. {\bf B} 540  587 (1999).

\bibitem{Smirnov}
 F.A. Smirnov, {\it Form-factors in completely integrable models of quantum field theory}, Singapore, World Scientific 1992.


\bibitem{Cardy}
J. Cardy, G. Mussardo Nucl. Phys. B {\bf 410} 451 (1993).

\bibitem{Fring}
A. Fring. G. Mussardo and P. Simonetti {\bf 393} 413, (1993). 

\bibitem{Mihalov}
Teoreticheskaya i Matematicheskaya Fizika, {\bf 30}, 303, (1977)

\bibitem{Bhattacharya}
T. Bhattacharya, J. Math. Phys. {\bf 46}, 012301 (2005).

\bibitem{LewensteinAhufinger12}
M. Lewenstein, A. Sanpera, and V. Ahufinger,
{\it Ultracold Atoms in Optical Lattices: Simulating quantum many-body systems}, Oxford University Press, (2012).

\bibitem{BlochDalibard12}
I. Bloch, J. Dalibard, and S. Nascimbene, Nat. Phys. {\bf 8}, 267 (2012). 

\bibitem{BlattRoss12}
R. Blatt and C. F. Ross, Nat. Phys. {\bf 8}, 277 (2012).

\bibitem{Cirac} 
J.I. Cirac, P. Maraner and J.K. Pachos, Phys. Rev. Lett.
\textbf{105}, 190403 (2010).

\bibitem{Barrett}
S. Barrett, et al., arXiv:1206.4988.

%


%



%
%
%

\bibitem{EIT}M. D. Lukin and A. Imamoglu Phys. Rev. Lett. \textbf{84}, 1419
(2000)

\bibitem{DPS1} Fleischhauer M and Lukin M D 2000 \textit{Phys. Rev. Lett.}
\textbf{84} 5094-7


\bibitem{DPS2} Bajcsy M \textit{et al} 2009 \textit{Phys. Rev. Lett.}
\textbf{102} 203902

\bibitem{Vetsch10} Vetsch E, Reitz D, Sague G, Schmidt R, Dawkins S T, and
Rauschenbeutel A 2010 \textit{Phys. Rev. Lett.} \textbf{104} 203603

%
%


\bibitem{Bajcsy11} M. Bajcsy, S. Hofferberth, T. Peyronel, V. Balic, Q.
Liang, A. S. Zibrov, V. Vuletic, and M. D. Lukin, Phys. Rev. A \textbf{83},
063830 (2011).

\bibitem{Chang08} D.E. Chang \emph{et al.}, Nat. Phys. \textbf{4}, 884 (2008).

\bibitem{Angelakis11} D.G. Angelakis \emph{et al.}, Phys. Rev. Lett.
\textbf{106}, 153601 (2011).



\bibitem{Tonks} L. Tonks, Phys. Rev. \textbf{50}, 955-963 (1936).

\bibitem{Girardeau} M. Girardeau, J. Math. Phys. (N.Y.) \textbf{1}, 516-523
(1960).




%


\end{thebibliography}
\end{document}